\documentstyle[pre,aps]{revtex}
\preprint{}

\begin{document}

\title{Diffusion of particles moving with constant speed}
\author{S.Anantha Ramakrishna \footnote{Electronic mail :
sar@rri.ernet.in} and N.Kumar \footnote{Electronic mail : nkumar@rri.ernet.in}}
\address{Raman Research Institute, Sadashivanagar, Bangalore-560 080, India}
\date{\today}
\maketitle
\begin{abstract}
	The propagation of light in a scattering medium is described as 
the motion of a special kind of a Brownian particle on which the fluctuating 
forces act only perpendicular to its velocity. This enforces strictly and
dynamically the 
constraint of constant speed of the photon in the medium. A Fokker-Planck 
equation is derived for the probability distribution in the phase space 
assuming the transverse fluctuating force to be a white noise. 
Analytic expressions for the moments of the 
displacement $\langle x^{n}\rangle$ along with an approximate expression 
for the marginal probability distribution function $P(x,t)$ are obtained. 
Exact numerical solutions for the phase space probability distribution 
for various 
geometries are presented. The results show that the velocity distribution 
randomizes in a time of about eight times the mean free time ($8t^*$) only 
after which the diffusion approximation becomes valid. This factor of eight
is a well known experimental
fact. A persistence exponent of $0.435 \pm 0.005$ is calculated for this 
process in two dimensions by studying the survival probability of the particle 
in a semi-infinite medium. The case of a stochastic amplifying medium is 
also discussed. \\ 

\end{abstract}
\section{Introduction}
    The propagation of light through a stochastic medium 
is traditionally described in the context of astrophysics 
by a Boltzmann transport
equation for the specific intensity $I(\vec{r},\vec{\Omega},t)$ in a
heuristic Radiative transfer theory \cite{chandra1}. However, since 
the general analytic
solutions are unknown, one resorts to the diffusion approximation which
can be shown to arise out of the Radiative transport equation in the limit of
large length scales $L >> l^{*}$, where $l^{*}$ is the transport mean free
path of light in the medium \cite{chandra1,ishimaru}. Recently there has
been considerable interest in the description of multiple light scattering
at small length scales $(L \sim l^{*})$ and small time scales ($t \sim
t^{*}$ where $t^{*}$ is the transport mean free time), both from the point
of fundamental physics \cite{kop} and from the point of medical
imaging, where the early arriving `snake' photons are used to image through
human tissues \cite{colak,alfano1}. It has been experimentally shown that
the diffusion approximation fails to describe phenomena at distances of $L
 < 8l^{*}$  \cite{alfano2}. Moreover the diffusion approximation which is strictly
a Wiener process for the spatial co-ordinates of a particle is physically
unrealistic. It holds in the limit of the mean free path $l^{*}
\rightarrow 0$ and the speed of propagation $c \rightarrow \infty$ while
keeping the diffusion coefficient $D_o =\frac{c l^{*}}{3}$ constant. Thus
the diffusion approximation neither accounts for a finite mean free path
nor for a finite and constant speed of the particle which is
charecteristic of light propagation in a stochastic medium. While approximately 
describing light as a particle the constancy of speed should be preserved 
at the very least. Hence it is of 
importance to develop better and alternative schemes to the diffusion
approximation and also address the difficult question of the process of
randomization of a directional beam in such media. \\

For a particle moving with fixed speed in a one-dimensional disordered medium,
 it has be shown that the probability distribution function $P(x,t)$ for the 
displacement  satisfies the telegrapher equation exactly. However, 
generalizations of the telegrapher equation to higher dimensions 
\cite{durian} have been shown not to yield better results than the diffusion 
approximation \cite{masoliver}.   
Recently there have been a few attempts to overcome the shortcomings
of the diffusion approximations and attack this problem using the concept
of photon paths. In Ref.\cite{feng}, a Monte-Carlo approach was used to
simulate photon paths and calculate their probabilities. An important
advance was made in Refs.\cite{feld1,feld2}, where the propagator for
photons in highly forward scattering media was expressed as a Feynman path
integral. However, this attempt has had only limited success in that it
was possible to calculate the probability distribution subject to the
constraint of constant photon speed only in the weaker (average) sense
{\it i.e.,} $\int^{t}_{0} \left[ (\frac{d\vec{r}}{dt})^{2} -c^{2} \right]
dt = 0$.  Moreover in
addressing  the backscattering from a semi-infinite
medium \cite{feld1} and reflection/transmission from a finite slab 
\cite{feld2}, the
absorbing boundary conditions have not been rigorously implemented and it
would be inappropriate to compare these to experimental data. 
It should be mentioned here that the Ornstein-Uhlenbeck (O-U) theory of
Brownian motion \cite{chandra2} would also be able to incorporate the
finiteness of the mean free path and a well defined root-mean-squared (rms)
velocity but assuming, of course a distribution of speeds. This process has
been compared  with Monte-Carlo simulations \cite{vgopal} and used to
explain the lowering of the effective diffusion coefficient measured in
pulse transmission experiments through thin slabs \cite{kop}. It can be 
shown that the
finite r.m.s speed defined by the fluctuation-dissipation theorem for the
O-U process is a stronger global constraint than the average speed
constraint imposed in Ref.\cite{feld1,feld2}.\\
    
    The next important step in describing these photon random walks with a
constant speed was undertaken in Ref.\cite{alfano3}, where the authors describe
this process 
as a non-Euclidean diffusion on the velocity sphere and
intuitively put down a kind of  a general Boltzmann equation for photons
in a highly forward scattering medium. The solution to this equation was
expressed as a path integral, which was then evaluated by a standard
cumulant decomposition \cite{kubo} truncated  after the second cumulant. This
yields a gaussian distribution similiar to the Ornstein-Uhlenbeck process.
More recently, an explicit derivation of the Feyman path integral
representation for the propagator of the radiative transfer equation has
been given \cite{miller}. Here it was again evaluated by truncating the
cumulant expansion after the second term. This was justified by declaring
that photons are massless and non-interacting. However, the imposition of
the speed constraint would not allow this gaussian approximation.\\

    In this paper,
we describe the light propagation in stochastic media as the motion of 
a kind of Brownian particle on which the fluctuating forces act only
perpendicular to the direction of its velocity. This is  effective
in strictly  and dynamically preserving the speed of the particle. 
This process is shown to
correspond to a diffusion in the angular co-ordinate in the velocity space
for a white noise disorder. Exact expressions for the moments of the
space variables are presented and the second cumulant approximation is
shown to yield a gaussian expression similiar to the traditional
Ornstein-Uhlenbeck theory of Brownian motion. An expression is
derived for the probability distribution for large force strengths which
preserves the light cone. The exact  Fokker-Planck equation for the probability
distribution is derived from the stochastic Langevin equations for a white
noise process. 
Numerical solutions of this equation are presented .
It is shown that the probability distribution in infinite media is
strongly forward peaked for short times and randomizes  only at times of
about $8 t^{*}$ to $10 t^{*}$. We have also solved numerically the
equation for a semi-infinite geometry and obtained the persistence exponent
of $0.435i\pm 0.005$ in 2-dimensions for this process. Solutions for a finite geometry 
are also given,
showing that the effective diffusion coefficient as measured in a pulse
transmission experiment through very thin slabs ($L \sim l^{*}$) 
would be lowered. The effect of light amplification 
in the slab is examined briefly.\\

\section{The modified Ornstein Uhlenbeck process}
	Light scattering in a stochastic medium is treated as a probabilistic 
process where each scattering event only changes the direction of the photon.
The wave nature and polarization effects are
ignored and light is treated as a  particle in a medium which exerts transverse
fluctuating forces on the particle. It should be remarked here that while
the actual disorder is maybe in space (quenched disorder), all current
treatments including ours, are in terms of a Brownian motion (temporal
disorder {\it i.e.,} a stochastic process). This is a valid 
approximation for incoherent transport in the weak scattering limit
($kl^{*} >> 1$ where $k=\frac{2 \pi}{\lambda}$, $\lambda$ being the
wavelength of light in the medium.
The equation for the motion of a randomly accelerated particle with the 
special condition that the random forces always act only perpendicular 
to the velocity can be written as, 
\begin{equation}
\ddot{\vec{r}}  =  \dot{\vec{r}} \times \vec{f}(t) \\
\end{equation}
This we term as the modified Ornstein-Uhlenbeck process. 
We will consider two dimesions for simplicity, and write 
\begin{eqnarray}
\ddot{x} &=& -f(t) \dot{y}   \label{SLE1} \\
\ddot{y} &=& f(t) \dot{x}	
\label{SLE2}
\end{eqnarray}
where the force term $f(t)$ is a random funtion of time. We will assume a delta-correlated force with gaussian distribution. {\it i.e.,} 
\begin{eqnarray}
\langle f(t) \rangle &=& 0 	\\
\langle f(t) f(t') \rangle &=&  \Gamma \delta (t-t') 	
\end{eqnarray}
and all higher moments of $f(t)$ being zero. This makes our treatment  most 
valid for a very dense collection of highly forward scattering 
weak anisotropic scatterers. This set of stochastic Langevin equations 
yield on integration a first constant of integration 
${\dot{x}}^{2} + {\dot{y}}^{2} = c^{2}$, where $c$ is the constant speed. 
So we can choose 
$ \dot{x} = c \cos {\theta (t)}$ and $\dot{y} = c \sin{\theta (t)}$ where 
$\theta (t)$ is some function of $t$. $\theta$ is recognised to be the angular
co-ordinate in the velocity space. Substituting these expressions back 
into equation(\ref{SLE1},\ref{SLE2}), we obtain $\dot{\theta} = f(t)$ or
\begin{equation}
\theta(t) - \theta_{0} = \int^{t}_{0} f(t) dt
\end{equation}
Hence $\theta(t)$ follows a Wiener process and we can write the probability for $\theta(t)$ as 
\begin{equation} 
P_{t} (\theta) = \left(\frac{1}{2 \pi \Gamma t}\right)^{\frac{1}{2}} \exp \left\{ -\frac{(\theta -\theta_{0})^{2}}{2 \Gamma t} \right\}
\label{angdif}
\end{equation}
This is the result for a diffusion in $\theta$ the angular co-ordinate in 
the velocity space, and we recognise this modified O-U process to be a 
random walk on the circle of radius $c$ in the velocity space. Constraining 
$\theta$ to the range $[0,2\pi]$, we get the marginal probability 
distribution for $\theta$. 
\begin{equation}
P_{t}(\theta) = \sum^{\infty}_{n=-\infty}\left(\frac{1}{2 \pi \Gamma t}\right)^{\frac{1}{2}}
\exp \left\{ -\frac{(\theta -\theta_{0} + 2 n \pi)^{2}}{2\Gamma t} \right\}
\end{equation} 
The value of $\theta_{0}$ can be conveniently chosen to be zero. \\

Now we will derive the probability distribution function in the phase space. Consider the
system of Stochastic Langevin equations
\begin{eqnarray}
\dot{x} = u \\
\dot{y} = v \\
\dot{u} = -f(t) v \\
\dot{v} = f(t) u
\end{eqnarray}
Let $\Pi(x,y,u,v)$ be the phase space density of points for the given system and
${\bf U}$ be the vector $(x,y,u,v)$. Now, 
$\Pi$ satisfies the stochastic Liouville equation. 
\begin{equation}
\frac{\partial \Pi}{\partial t} + \nabla_{\bf U} \cdot \left(\dot{\bf U} \Pi\right)  = 0
\end{equation}
 where $\nabla_{\bf U} = ( \frac{\partial}{\partial x}, \frac{\partial}{\partial
 y}, \frac{\partial}{\partial u}, \frac{\partial}{\partial v})$.
 Substituting for ${\bf U}$ and averaging over all possible configurations
of disorder, by the van Kampen Lemma \cite{vankampen}, 
the probability distribution $P(x,y,u,v) = \langle \Pi (x,y,u,v) 
\rangle$ and satisfies,
\begin{equation}
\frac{\partial P}{\partial t} + u \frac{\partial P}{ \partial x} + 
v \frac{\partial P}{\partial y}
- v \frac{\partial}{\partial u} \langle f(t) \Pi \rangle + u \frac{\partial}
{\partial  v}
\langle f(t) \Pi \rangle = 0
\end{equation}
By the Novikov theorem \cite{novikov} for a white noise process $f(t)$,
\begin{eqnarray}
\langle f(t) \Pi[f(t)] \rangle & = & \frac{\Gamma}{2} \langle ~\frac{\delta \Pi[
f]}
{\delta f(t)} ~\rangle  \nonumber   \\
 & = &  \frac{\Gamma}{2}\left( v \frac{\partial P}{\partial u} - u \frac{\partial
P}{\partial v} \right)
\end{eqnarray}
Using the above, we obtain for P(x,y,u,v) the differential equation
\begin{equation}
\frac{\partial P}{\partial t} + u \frac{\partial P}{ \partial x} + v 
\frac{\partial P}{\partial y}
= \frac{\Gamma}{2} \left( u \frac{\partial  }{\partial v} - v \frac{\partial }
{\partial u}
\right)^{2} P
\end{equation}
Now expressing u and v in terms of the angular co-ordinate $\theta$, we 
finally get
\begin{equation}
\frac{\partial P}{\partial t} + c~\cos\theta \frac{\partial P}{ \partial x} + c~
\sin\theta
 \frac{\partial P}{\partial y} = 2 \Gamma \frac{\partial^{2} P}{\partial \theta^
{2}}
\label{ned}
\end{equation}
This
differential equation explicitly preserves the constancy of the speed of
the photon. This Fokker-Planck equation is the same equation (in two dimensions) which
was written down in Ref.\cite{alfano3}. 
It is rigorously proved therein that  this has a path integral solution and the
two approaches are equivalent. It appears that this equation has solutions 
in terms of Mathieu functions. However we have not been able to analytically 
solve the equation.\\ 

	The  moments of the displacements can however be calculated 
analytically. The
displacements can be written in terms of $\theta$ as 
\begin{eqnarray}
\label{xy}
x- x_{0} & = & \int^{t}_{0} c~ \cos \theta d\theta   \\
y- y_{0} & = & \int^{t}_{0} c~ \sin \theta d\theta   
\end{eqnarray}
Using these and a gaussian distribution for $f(t)$, we get
\begin{eqnarray}
\langle x- x_{0} \rangle & = & \frac{2c}{\Gamma}\cos\theta_{0}\left( 1- e^{-\frac{\Gamma t}{2}} \right)  \label{mom1}\\
\langle y- y_{0} \rangle & = & \frac{2c}{\Gamma}\sin\theta_{0}\left( 1- e^{-\frac{\Gamma t}{2}} 
\right) \\
\langle (x- x_{0})^{2} \rangle & = & c^{2}\left[ \frac{2t}{\Gamma} - \frac{2}{3} \left(\frac{2}{\Gamma}\right)^{2}\left( 1- e^{-\frac{\Gamma t}{2}} \right) - \frac{1}{12}\left(\frac{2}{\Gamma}\right)^{2}\left( 1- e^{-\Gamma t} \right) \right] 	\label{mom2}\\
\langle (y- y_{0})^{2} \rangle & = & c^{2}\left[ \frac{2t}{\Gamma} - \frac{4}{3} \left(\frac{2}
{\Gamma}\right)^{2}\left( 1- e^{-\frac{\Gamma t}{2}} \right) + \frac{1}{12}\left(\frac{2}{\Gamma}
\right)^{2}\left( 1- e^{-\Gamma t} \right) \right]    \\
\langle (x-x_{0})(y- y_{0}) \rangle & = & 0 	
\end{eqnarray}
This reproduces the result of the traditional Ornstein-Uhlenbeck process in that the 
first moment saturates at a mean free path $l^{*}$ and  the second moment increases 
linearly with time at long times ($\Gamma t/2 \gg 1$).  For short times ($\Gamma t/2 \ll 1$),
the longitudinal spread $\langle \Delta x^{2} \rangle \sim t^{2}$ and the lateral spread 
 $\langle \Delta y^{2} \rangle \sim t^{3}$ which are considerably slower than the 
diffusive linear behaviour.  From these relations, we identify the mean free time
$t^{*}$ to be $\frac{2}{\Gamma}$ and the transport mean free path $l^{*} = ct^{*}$. The 
diffusion coefficient is identified as the coefficient of the linear term 
of the second moment {\it i.e.,} $\frac{c^{2}}{\Gamma}$  \\

It is of interest to note that an analytic expression for moments of 
all orders for the displacements can be obtained. This expression is given in 
the Appendix. The marginal probability distribution function 
$P(x,y,t;x_{0},y_{0},0)$ can be written in terms of a cumulant 
expansion (See the Appendix).  
Truncation  of the cumulant series 
after the second term yields the result of Ref.\cite{alfano3} for the 
probability distribution.
\begin{eqnarray}
P(x,y,t;x_{0},y_{0},0) & = & \frac{1}{2\pi det(M)} \exp\left\{-\frac{M_{ij}^{-1}}{2}
(\vec{r}-\vec{r_{0}}-\vec{a})_{i}(\vec{r}-\vec{r_{0}}-\vec{a})_{j} \right\} 	\\
\vec{a} & = & \frac{2c}{\Gamma}(1-e^{-\Gamma t/2})(\cos\theta_{0},\sin\theta_{0}) \nonumber \\
M_{ij} & = & \langle ~(\vec{r}-\vec{r_{0}})_{i} (\vec{r}-\vec{r_{0}})_{j}~\rangle -
\langle ~(\vec{r}-\vec{r_{0}})_{i}~\rangle \langle~(\vec{r}-\vec{r_{0}})_{j}~ \rangle  
\nonumber 
\end{eqnarray}
The distribution is gaussian in this approximation and similiar to the distribution for the traditional O-U process \cite{chandra2}. Thus it does not exactly preserve the light cone and would appear to constrain the speed only in an average sense. Higher cumulants would be required to describe this feature of fixed speed.\\
	
	An approximate solution which preserves the light cone can be 
obtained under the assumption that $\theta$ is completely randomized in
time $t^{*}$, so that $\theta$ has a uniform distribution over $[0,2\pi]$. 
This can be justified in the limit of large force strength ($\Gamma$), 
when the scattering events change the momentum by a large amount. 
Now the time can be discretized on this time-scale and the probability 
distribution can be written as
\begin{equation}
P(x,t;x_{0},0) = \frac{1}{2\pi} \int_{-\infty}^{\infty} d\omega ~e^{i\omega (x-x_{0})}
\langle ~ \exp [ -i\omega c \sum_{j=1}^{n} \cos\theta_{j} t^{*} ] ~ \rangle
\end{equation}
where we have used that at $t=0$, the angle $\theta$ was uniformly distributed. To evaluate 
the average, we will use the fact that in this approximation, each $\theta_{j}$ is independent
of all others, giving
\begin{equation}
\langle ~ \exp [ -i\omega c \sum_{j=1}^{n} \cos\theta_{j} t^{*} ] ~ \rangle  
  =  \left[ J_{0}(\omega c t^{*})\right]^{n}
\end{equation}
where $J_{0}$ is the ordinary Bessel's function of order zero. 
Using $n=t/t^{*}$, we have 
\begin{equation}
P(x,t;x_{0},0) = \frac{1}{2\pi} \int_{-\infty}^{\infty} d\omega e^{i\omega (x-x_{0})}
\left[ J_{0}(\omega c t^{*})\right]^{t/t^{*}}
\end{equation}
Using the fact that the Fourier transform of $J_{0}(\omega c t^{*})$ is 
zero for $\left| x-x_{0}\right| > c t^{*} $ and the fact that 
$P(x,t,x_{0},0)$ is an $n^{th}$ convolution of  $J_{0}(\omega c t^{*})$, 
it is seen that $P(x,t,x_{0},0)$ is zero for $\left| x-x_{0
}\right| > n c t^{*} = c t $. Thus the light cone is preserved. In fig-1, we 
plot the $P(x,t,x_{0},0)$  obtained by numerical evaluation for different 
times. It is seen that for $t=t^{*}$, the probability is accumulated at 
the light front, and all the curves show a cut-off at $\left| x-x_{0}
\right| = c t$. At long times, using the Laplace approximation, 
we have (for large $n$);
\begin{eqnarray}
\left[ J_{0}(\omega c t^{*})\right]^{n} & \simeq & \exp\left\{ -\frac{c^{2}{t^{*}}^{2}{\omega}^{2} n}{4}\right\} \nonumber	\\
P(x,t;x_{0},0) & \simeq & \left( \frac{1}{\pi c^{2} t^{*} t}\right)^{1/2} \exp \left\{ -
\frac{(x-x_{0})^{2}}{c^{2} t^{*} t} \right\}
\end{eqnarray}
Thus we recover the diffusion limit at long times.\\

\section{Numerical Solutions and Results}
	In this section , numerical solutions for the differential equation 
(\ref{ned})  are presented. The particle is released in the $x-y$ plane at the 
origin (generally) along an initial direction $\theta_{0}$. Here $\theta$ is the angle
made by the velocity vector with the $x$-axis. Let us first further simplify 
by assuming invarience with respect to $y$ {\it i.e.,} we have a line source along 
the $y$ axis. Then the derivative with respect to $y$ drops out and and we 
have a partial
differential equation in three variables. This is essentially a parabolic 
equation with an advective term. To numerically propagate the probability 
distribution in time, we use an alternating direction implicit -explicit 
method \cite{ames} for $x$ and $\theta$. A local von Neumann stability 
analysis  \cite{ames} shows that this differencing scheme is unconditionally 
stable. The initial condition is a $\delta$-function at $x=0, \theta = 0$ 
which is approximated by a sharp gaussian for numerical purposes. 
For infinite media, the boundary condition $P(x,t) = 0$ for 
$ \left| x \right| ~>~ c t$ is used. For a semi-infinite medium 
$-\infty < x < L$ with an absorbing boundary at $x=L$, the appropriate 
boundary condition is given by
$P(L,\theta,t;x_{0},\theta_{0},0) = 0 $ for $-\pi < \theta < -\pi/2$ and  
$\pi/2 < \theta < \pi$, corresponding to no flux entering the medium from 
free space. Also, we can write the Fokker-Plank equation in the form of 
a continuity equation.
\begin{eqnarray}
& &  \frac{\partial P}{\partial t}  + \nabla \cdot \vec{\jmath} = 0   \\
& &  \nabla  = \hat{e}_{x} \frac{\partial}{\partial x} + \hat{e}_{\theta}
 \frac{\partial}{\partial \theta}       \nonumber       \\
& & \vec{\jmath}  =  \hat{e}_{x} \cos\theta P
 + \hat{e}_{\theta} \Gamma \frac{\partial P}{\partial \theta}   \nonumber
\end{eqnarray}
Since $\Gamma = 0$ outside the medium, we can conclude that the current 
density $ \vec{\jmath} $ in
the real ($x$) space is conserved across the boundary in the forward direction
$(-\pi /2 < \theta < \pi /2 )$ while the current density in the velocity ($\theta$)
space is not conserved. This explains why the output flux at the boundary is 
proportional to the value of the probability distribution function at the 
boundary itself ( rather than the space derivative of the 
probability distribution 
$(\frac{\partial P}{\partial x})$ given by Fick's law ) as observed in
experiments \cite{alfano4}. For a finite slab we use a similiar boundary
condition at the other boundary.\\

    In Fig-2, we show the probability distributions in an infinite medium
with the initial condition, $P(x,\theta , t=0) = \delta(x) \delta(\theta)$.
It is clearly seen that the probability distribution for times upto $5t^{*}$
is peaked in the forward direction $\theta \sim 0$ for $x>0$, with a tail
in the backward direction $(\theta \sim \pm \pi)$ at $x<0$. There is also a
clear cut off at $\left| x \right| = c t$, which is prominently noticeable
for  positive $x$. The small amount of tailing arises from the finite
width of the gaussian by which the $\delta$-function was approximated. One
can also note that the probability distribution becomes almost flat along the
$\theta$-axis only at times of about eight times the mean free time
($8t^{*}$).  In Fig-3, the first and second moments of the $x$ co-ordinate
are shown. The solid lines show the analytical results of
equation(\ref{mom1} and \ref{mom2})  and the symbols($\circ$) 
represent the results of the
numerical solutions. Excellent agreement is found between them. In Fig-4,
we show the marginal probability distribution for $x$ {\it i.e.,}
$P(x,t;x_{0},\theta_{0},0) = \int_{-\pi}^{\pi} d\theta
P(x,\theta,t;x_{0},\theta_{0},0)$. At short times ($t\simeq 3t^{*}$), there is
a clear ballistic peak, separate from the more randomized tail. The probability
distribution for these times is also clearly forward-peaked. One can also note
that the probability distribution randomizes and becomes almost gaussian,
centred at $x \sim l^{*}$ only at times $t\geq 8t^{*}$. As noted above,
this is also the time by which the angular coordinate $\theta$ randomizes.
This is when the diffusion approximation becomes valid. This can be
understood by noting that, by equation(\ref{angdif}), the time required for
$P_{t}(\theta)$ to attain an angular width of $2\pi$ is $T$ where $T$
is given by $\langle\Delta \theta^{2} \rangle = (2\pi )^{2} \sim 2\Gamma T$.
This yields (using $\Gamma /2 = t^{*}$) a value  of $T= \pi^{2} t^{*}
\simeq 10t^{*}$ for the randomization time. 
Thus we now have a clear picture of the
reason for this long known experimental fact \cite{alfano2}. This forward
peaked behaviour at short times also illustrates the deficiency of the
second cumulant approximation where the probability distribution is a
Gaussian and symmetric about the first moment. Higher cumulants are
clearly required to describe these asymmetric features.  \\

The probability distribution  functions for a semi-infinite medium  are shown in
Fig-5. Here the particle is released at the origin inside the random
medium and the initial direction is towards the boundary 
(in this case at $x=4l^{*}$)
For times lesser than $4t^{*}$, there is no difference in the probability
distribution  from the case of the infinite medium. This is because the
wave front has not propagated upto the boundary and the effect of the
boundary is not felt. This is to be contrasted with the diffusion
approximation where the effect of the boundary is felt everywhere
simultaneously and causality is violated. At long times the probability
distributions attain a typical shape with a long tail at negative $x$ within the medium and a
sharp cut-off at the boundary.  In Fig-6, we show the marginal probability distribution for $x$
 {\it i.e.,}
$P(x,t;x_{0},\theta_{0},0) = \int_{-\pi}^{\pi} d\theta
P(x,\theta,t;x_{0},\theta_{0},0)$. The value of
$P(x,t;x_{0},\theta_{0},0)$ is finite at the boundary and zero
outside. As seen in Fig-6b, if the points near the boundary are linearly
extrapolated outside the boundary, they all roughly cross the x-axis at
about $0.7l^{*}$ which is the value of the extrapolation length used in
the diffusion approximation \cite{morse}. In Fig-7, the surviving probability inside
the medium $P_{s} = \int dx \int d\theta P(x,\theta,t;x_{0},\theta_{0},0)$
is plotted with time. For long times, this quantity should scale as
$t^{-\vartheta}$ where $\vartheta$ is the persistence exponent for this process
\cite{majumdar}. We have performed these calculations for several
source-boundary distances and obtained a value of $0.435 \pm 0.005$ as the
persistence exponent for this process in two dimensions. \\

Finally we present solutions for a finite slab with absorbing boundaries at $x=\pm L$. 
The particle is released from the origin at $t=0$ along the positive $x$ direction. 
Fig-8a shows the first and second moments of the probability with time in a 
thin slab of thickness $2l^{*}$. The first and second 
moments initially increase as in an unbounded medium until the photon-front hits the boundary 
and dip before increasing again and saturating  at an almost constant value. 
The dips occur because just after the ballistic and near ballistic 
components exit the slab, only the photons which are effectively moving in 
the opposite directions are left behind. In fact, the first moment is seen to
become negative, implying that the net transport is in the 
backward direction for some time. The dip in the second moment implies  
that the photon cloud  
is effectively expanding at a slower rate. This would cause a lowered 'effective 
diffusion coefficient' to be measured in a pulse transmission measurement. 
This reinforces the conclusions 
reached in Ref.\cite{vgopal} based on Monte-carlo simulations and explains 
the experimental results of Ref.\cite{kop} on a more rigorous footing.
Fig.8b shows the survival probability for the case of a finite slab. This  
decays considerably faster than the in case of the semi-infinite slab, 
though at early times ($t \sim t^{*}$) the decay rates are comparable. 
The initial rates of decay are comparable because of the forward peaked nature of the probability distribution 
at early times, when the effect of the boundary at the back is hardly felt. 
This is to be compared with the mirror-image solution in 
 the diffusion approximation, 
where equal weightage is given to both 
boundaries at all times.   \\
	
	Finally we turn to the case of an amplifying stochastic medium. The
effect of medium gain can be incorporated straight forwardly by 
noting that in our treatment, the time of exit from the 
slab directly translates into path-length traversed within the medium 
because speed is kept absolutely fixed. In the presence of amplification 
in the medium therefore, the net gain is directly proportional to the 
time. Thus the output flux at the boundary in a given direction is simply 
$P(L,\theta,t) \cos\theta \exp (\alpha t)$, where $\alpha$ is the gain 
coefficient in the medium. It is thus simple to obtain a picture of 
amplified emission from such a medium. In Fig-9, we show the total light 
emitted by slab with boundaries at $x=\pm 2l^{*}$ for several amplification 
factors. The photon is released from the origin in the positive $x$ -
direction. For large times, the output increases exponentially because of 
the presence of a exponential gain in the  medium with no saturation. It is 
seen that the ballistic part is only slightly amplified while the output in 
the tail regions  are increased considerably. To obtain a more realistic 
picture of lasing in
random media \cite{lawandy,hema} however, one would have to consider the lasing 
level population depletion and saturation effects. 

\section*{acknowledgement} 
SAR would like to thank Prof. Rajaram Nityananda for very helpful discussions.

\section*{Appendix : Expression for the moments $\langle x^{n} \rangle$}
The $n^{th}$ order moment is given by
\begin{equation}
\langle ( x - x_{0})^{n} \rangle = c^{n} \int^{t}_{0} \int^{t}_{0} \cdots \int^{t}_{0} 
dt_{n} dt_{n-1} \cdots dt_{1} \langle \cos \theta(t_{1}) ~\cos \theta(t_{2}) \cdots
\cos \theta(t_{n}) \rangle
\end{equation} 
Writing $\theta(t_{i})$ as $\theta_{i}$, the quantity within the angular brackets can be expressed as follows,
\begin{eqnarray}
\langle \cos \theta_{1} ~\cos \theta_{2} \cdots
\cos \theta_{n} \rangle & = & 2^{-n} \langle ~(e^{i\theta_{1}} + e^{-i\theta_{1}}) 
(e^{i\theta_{2}} + e^{-i\theta_{2}}) \cdots (e^{i\theta_{n}} + e^{-i\theta_{n}}) ~\rangle   \nonumber \\
 & = & 2^{-n} \sum_{\stackrel{\sigma_{1}, \sigma_{2} \cdots \sigma_{n}}{ \sigma_{i} = \pm 1}} \langle ~~\exp [ i\sum_{j=1}^{n} \sigma_{j} \theta_{j} ] ~~\rangle
\end{eqnarray}
This can be expressed as a path integral using a gaussian distribution for $f(t)$.
\begin{eqnarray}
\langle ~~\exp [ i\sum_{j=1}^{n} \sigma_{j} \theta_{j} ] ~~\rangle & = & \int {\cal D}[f(t)] 
\exp\left\{-\int^{t}_{0} \left[ \frac{f^{2}(t')}{2 \Gamma} + i\sum_{j=1}^{n} \sigma_{j} 
f(t')\right] dt'\right\}   \nonumber 	\\
  & = & exp \left\{-\frac{\Gamma}{2} \sum_{k=1}^{n} \left(\sum_{j=k}^{n} \sigma_{j} \right)^{2} (t_{k} - r_{k-1} ) \right\}
\end{eqnarray}
where $t_{0} = 0$ and we assumed a time ordering of $t_{1} < t_{2} < ~\cdots~ < t_{n}$. Thus,
\begin{equation}
\langle ( x - x_{0})^{n} \rangle = c^{n} (n!) 2^{-n} \int^{t}_{0} dt_{n} \int^{t_{n}}_{0} dt_{n-1} \cdots \int^{t_{2}}_{0} dt_{1} \sum_{\stackrel{\sigma_{1}, \sigma_{2} \cdots \sigma_{n}}{
\sigma_{i} = \pm 1}}\exp \left\{-\frac{\Gamma}{2} \sum_{k=1}^{n} \left(\sum_{j=k}^{n} \sigma_{
j} \right)^{2} (t_{k} - r_{k-1} ) \right\}
\end{equation}
A similiar expression can be obtained for the $\langle (y-y_{0})^{n} \rangle$
by noting that $ \sin\theta = \cos (\pi /2 - \theta)$.\\

	Now we can obtain the joint probability distribution of $x$ and $ y$ as
\begin{equation}
P(x,y,t;x_{0},y_{0},0) = \langle ~\delta(x-x_{0}-c \int_{0}^{t} \cos\theta(t') dt')~ \delta
(y-y_{0}-c \int_{0}^{t} \sin\theta(t') dt')~ \rangle
\end{equation}
Expressing the $\delta$-functions in terms of the Fourier transforms, 
\begin{equation}
P(x,y,t;x_{0},y_{0},0) =  \left( \frac{1}{2\pi} \right)^{2} \int_{-\infty}^{\infty}
d\omega_{x}\int_{-\infty}^{\infty}d\omega_{y} ~e^{i[\omega_{x}(x-x_{0})+\omega_y(y-y_{0})]}
\langle ~\exp \left[ -ic\int_{0}^{t} [\omega_{x}\cos\theta(t') + \omega_{y}\sin\theta(t')] dt'
\right] ~ \rangle
\end{equation}
This statistical average can be evaluated by a cumulant expansion \cite{kubo} 
and since we have an expression for moments of all orders, we can in 
principle evaluate the cumulant expansion to any desired order.

\pagebreak

\newpage
\begin{center} {\bf FIGURE CAPTIONS}\\
\end{center}

\noindent Figure-1 : The marginal probability distributions $P(x,t;x_0,0)$
predicted by the approximate solution given by  equation(28) at different times
indicated in the figure. There is a clear cut-off at the light front and 
initially the probability accumulates at the light front (for $t =t^{*}$).\\

\noindent Figure-2 : The Probability distributions in the phase space 
of a particle in an infinite medium at
different times obtained by numerically propagating equation(17). The particle
is released at $x=0$ along the positive $x$ direction($\theta = 0$) at  $t=0$.
The probability distribution is clearly forward peaked and and becomes almost
flat along the $\theta$-axis only at times of about $8t^{*}$.\\

\noindent Figure-3 : The first and second moments of the displacement for the 
probability distribution of a particle in an infinite medium. The solid lines 
show the anlytical result of equation(20) and equation(22) while the symbol (
$\circ$ )
show the result obtained from the numerical solutions.\\

\noindent Figure-4 : The marginal probability distribution $P(x,t;x_0,\theta_0,0) = \int_{-\pi}^{\pi} P(x,\theta,t;x_0\theta_0,0) d \theta$ at different times.
The marginal probability distribution becomes almost a Gaussian at times of $8t^{*}$.\\

\noindent Figure-5 : The Probability distributions in the phase space 
of a particle in a semi-infinite medium at
different times. The particle  
is released at $x=0$ along the positive $x$ direction($\theta = 0$) at  $t=0$.
The absorbing boundary is located at $4l^{*}$. The probability distribution  is 
zero in the range $-\pi < \theta < -\pi/2$ and $\pi/2 < \theta < \pi$ at 
the boundary implying there is no incoming flux into the medium.\\

\noindent Figure-6 : The marginal probability distributions in a semi-infinite medium with an absorbing boundary at $x = 4l^{*}$. The plot on the right shows an expanded view of the distributions near the boundary. The solid straight lines
are the linear extrapolations of the behaviour near the boundary. All of them 
are seen to cross the $x$-axis roughly at $0.7l^{*}$  outside the boundary.\\

\noindent Figure-7 : The surviving probability of the particle inside the 
semi-infinite medium for an absorbing boundary at $4l^{*} (\circ)$ and $2l^{*} 
(*)$. The persistence exponent $\vartheta$ is obtained from the long
time behaviour of the survival probability. The lines ($\cdots$) and ($-\cdot-$)
show the linear fits and give a persistence exponent of 0.4309 and 0.4364 respectively.\\

\noindent Figure-8 : The first moments of the displacement of a particle 
in a finite slab of thickness $2l^{*}$ (left plot). The right plot shows 
the survival probability in a semi-infinite medium and a finite slab. The 
distance between the point where the particle is released and the boundary
is same in both case. ($2l^{*}$).\\

\noindent Figure-9 : Total light emitted (from both sides) by a disordered slab with
amplification for different values of the gain 
coefficient $A$ in the medium. The output increases exponentially at long times.\\

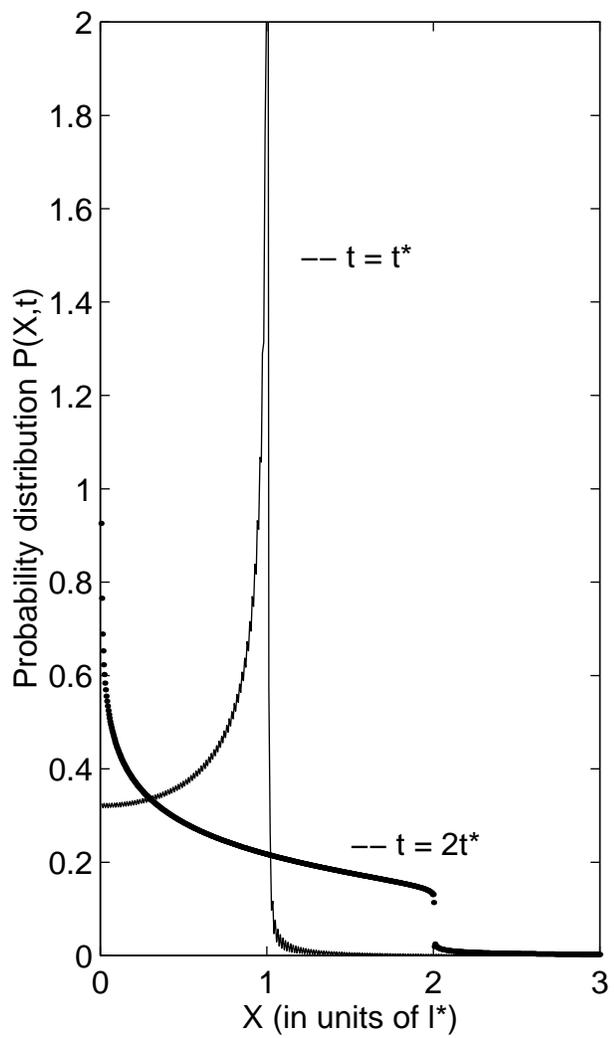
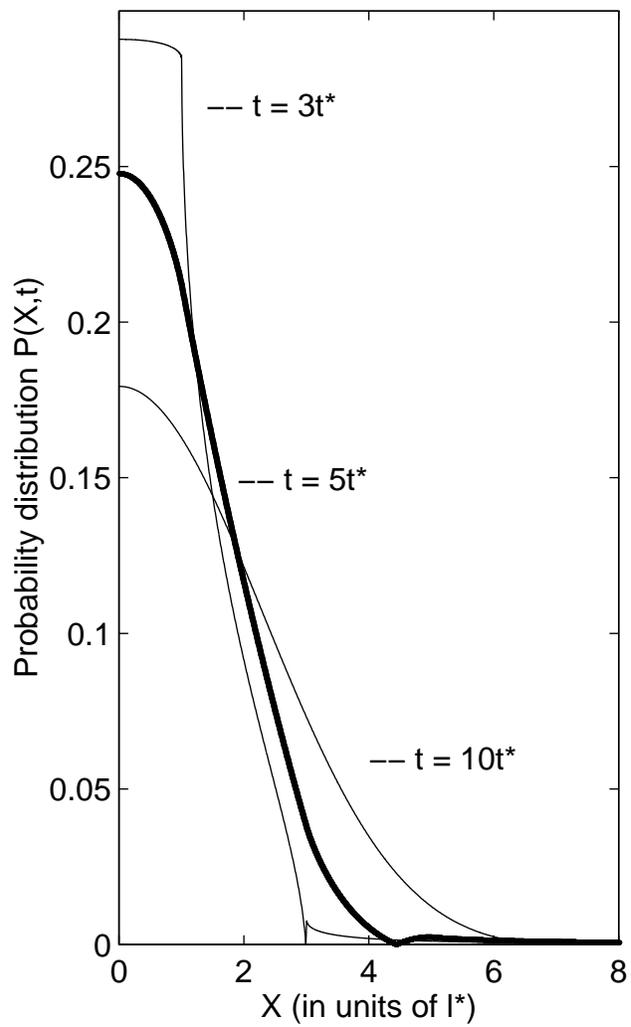
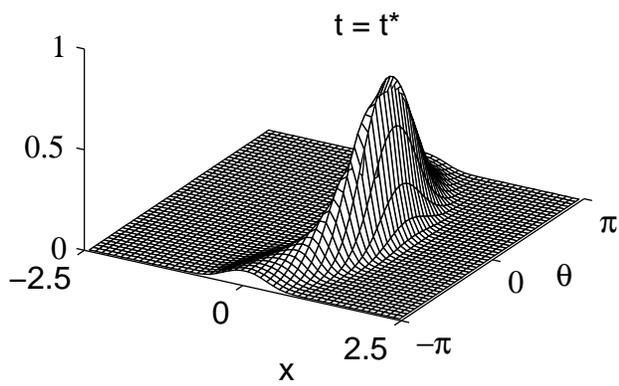
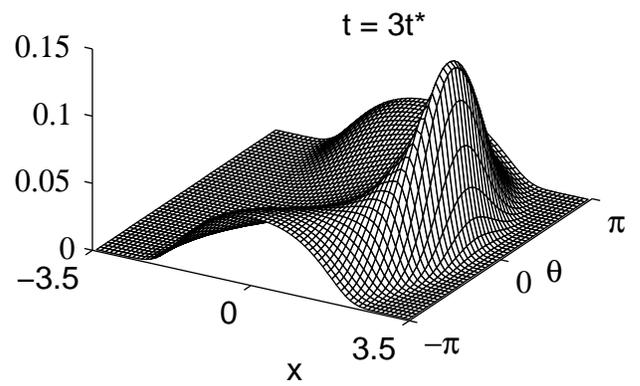
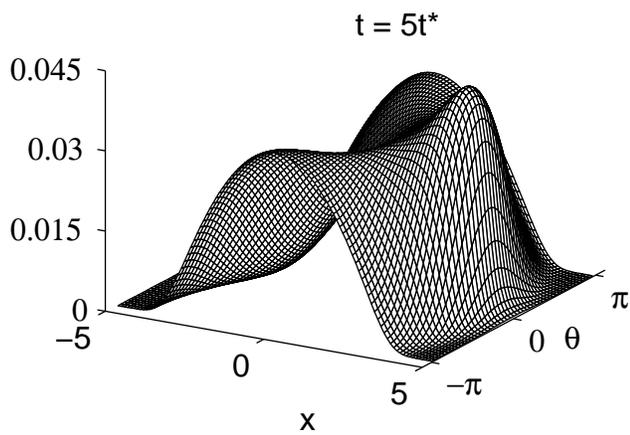
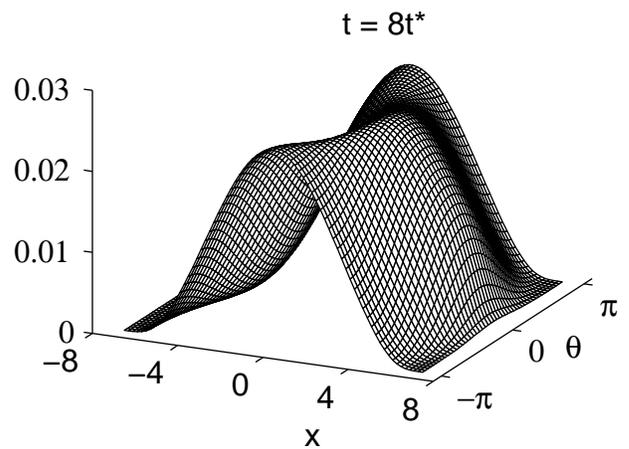
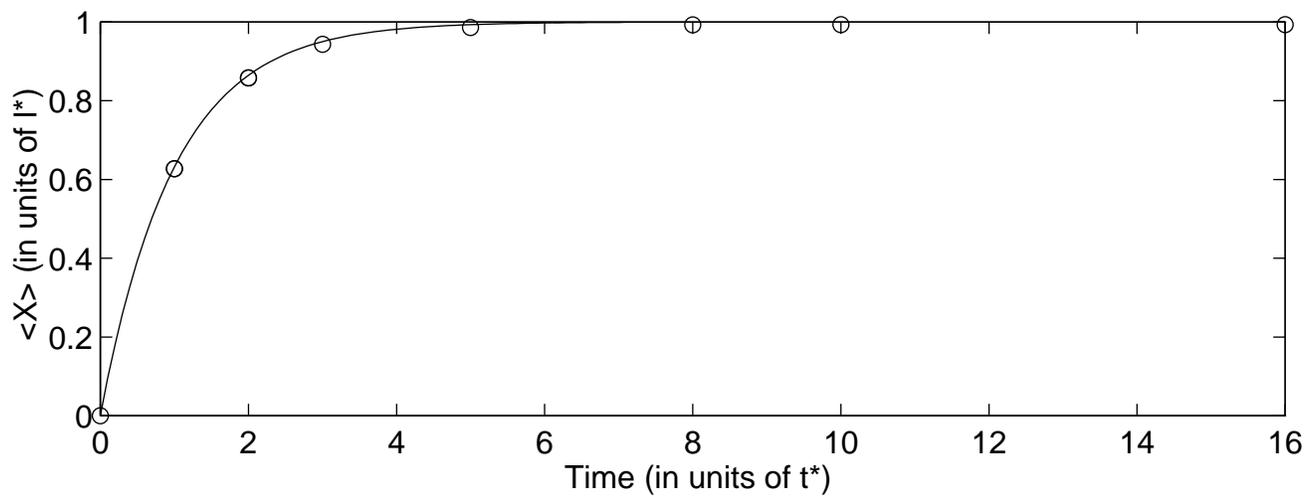
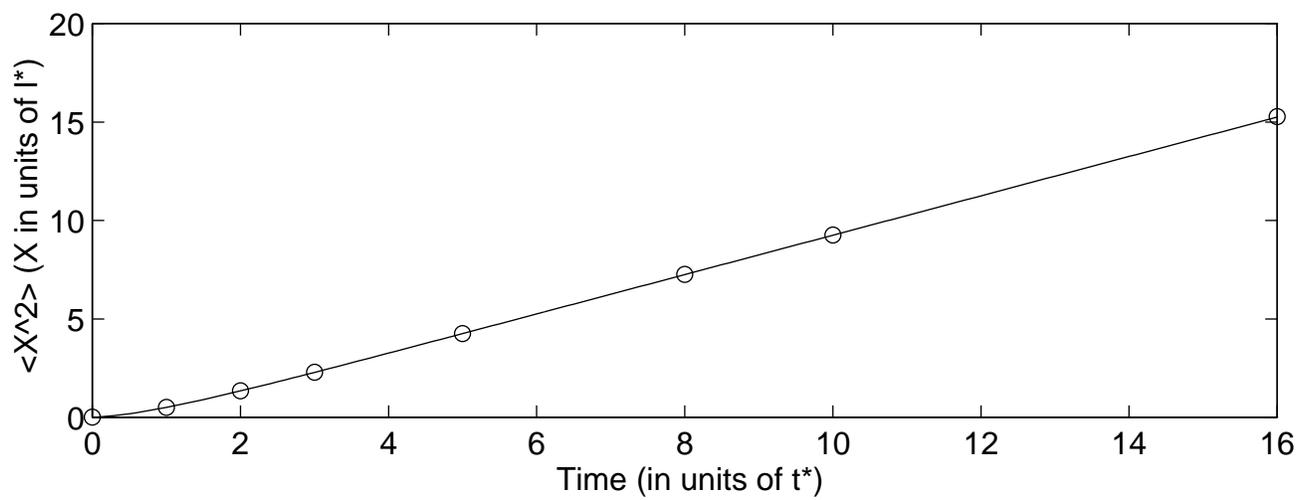
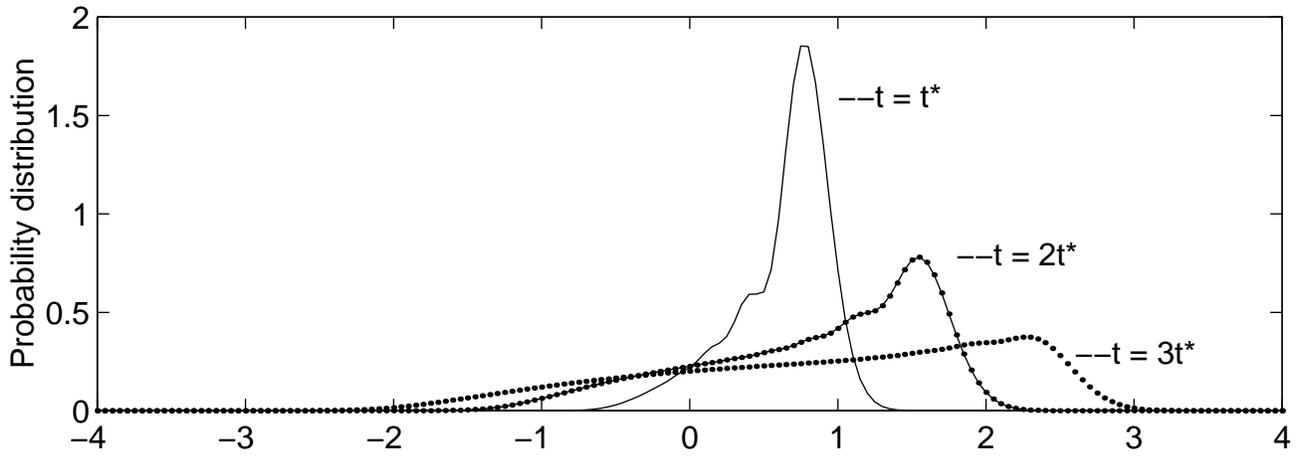
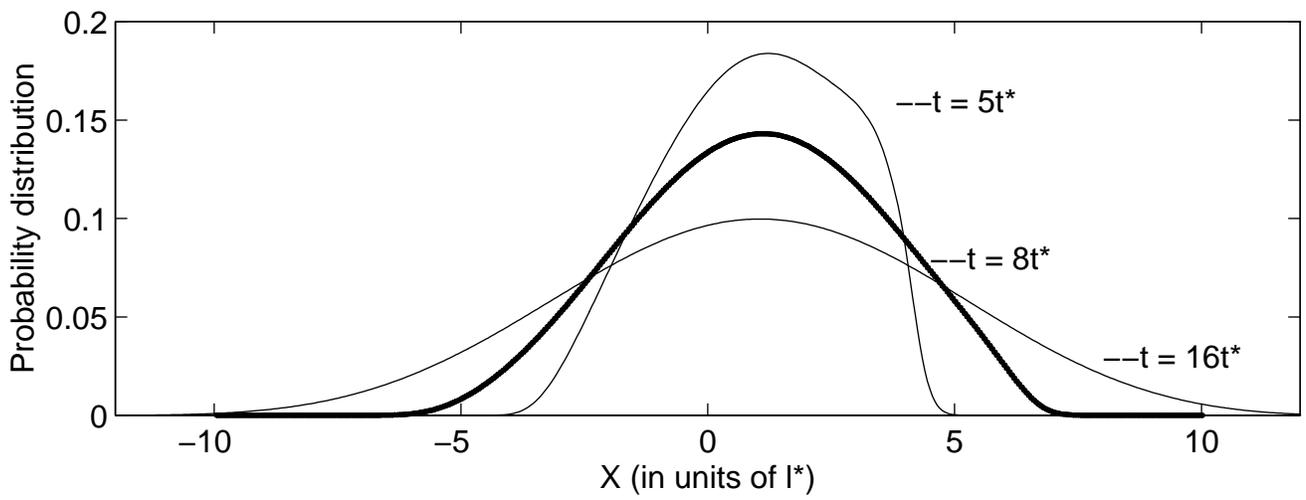
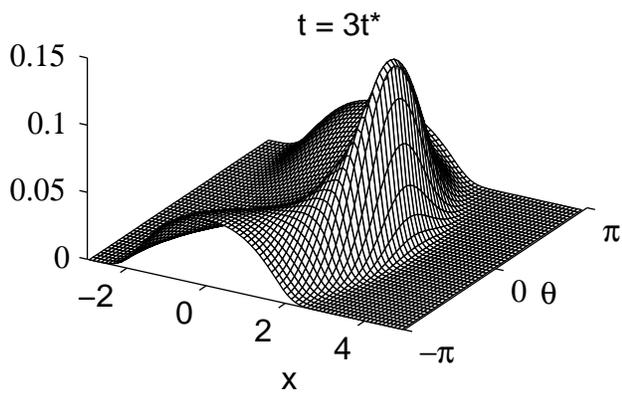
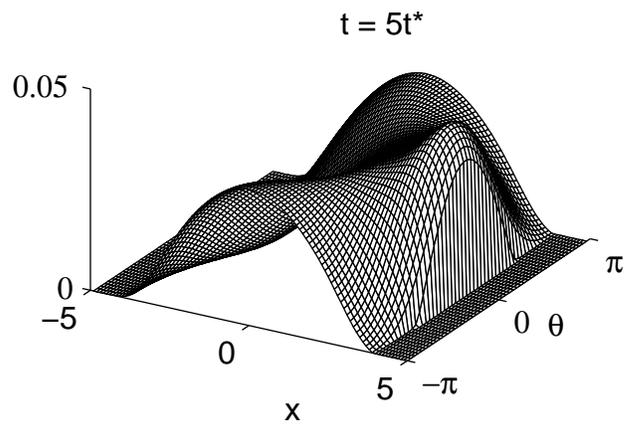
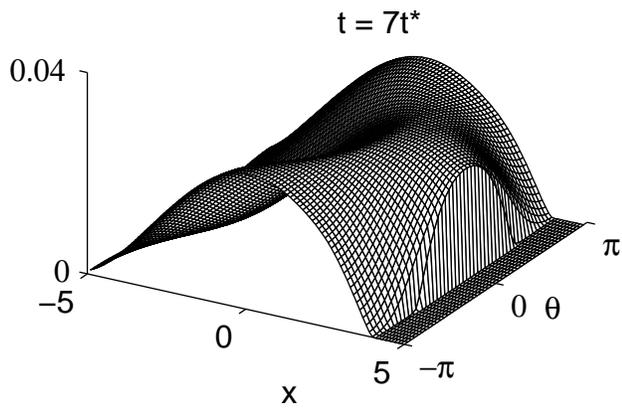
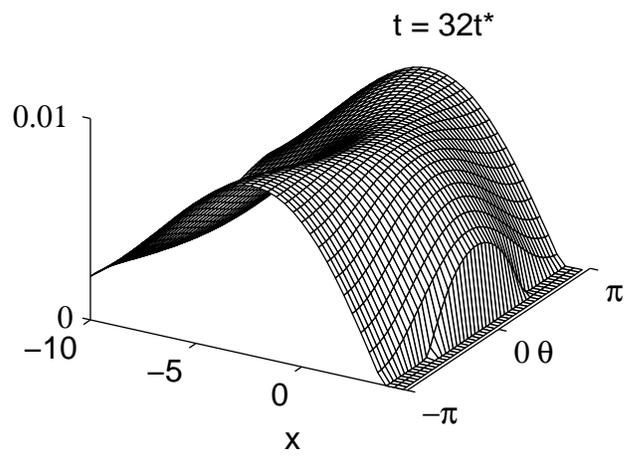
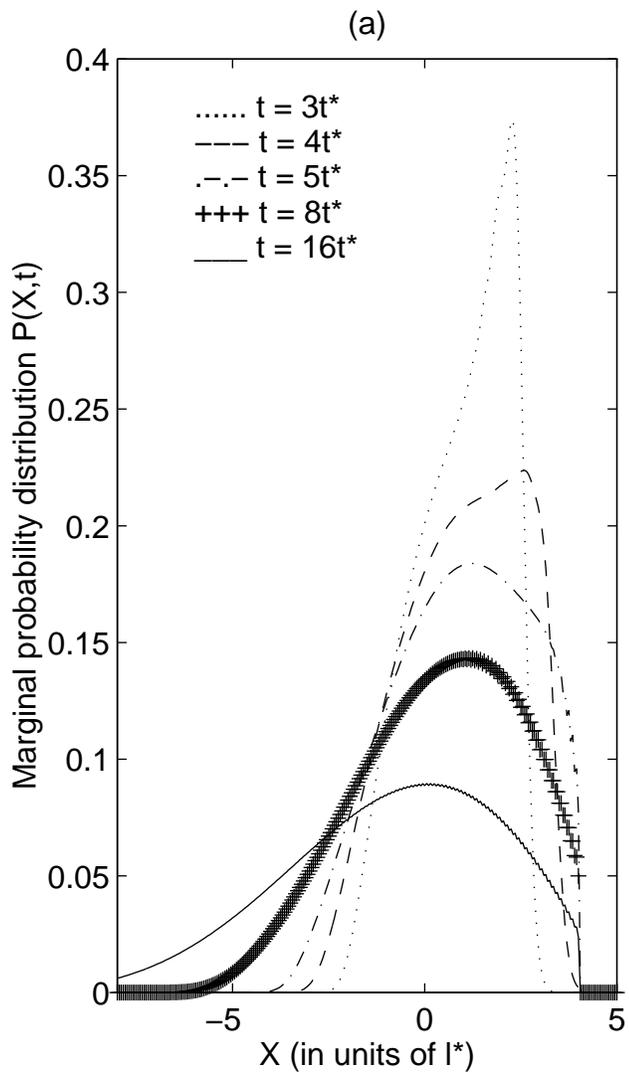
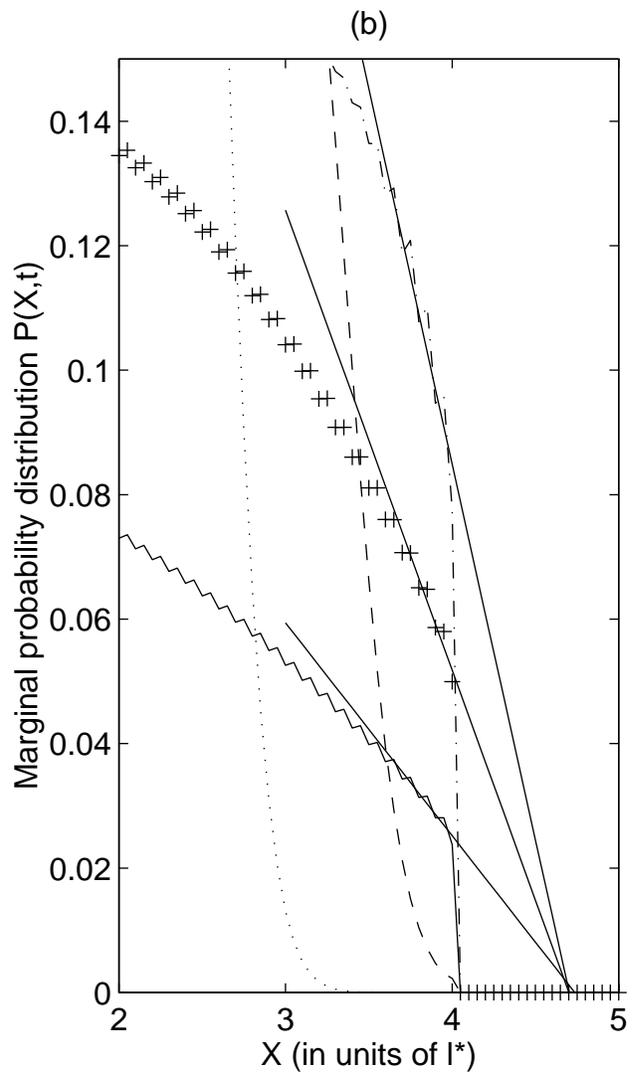
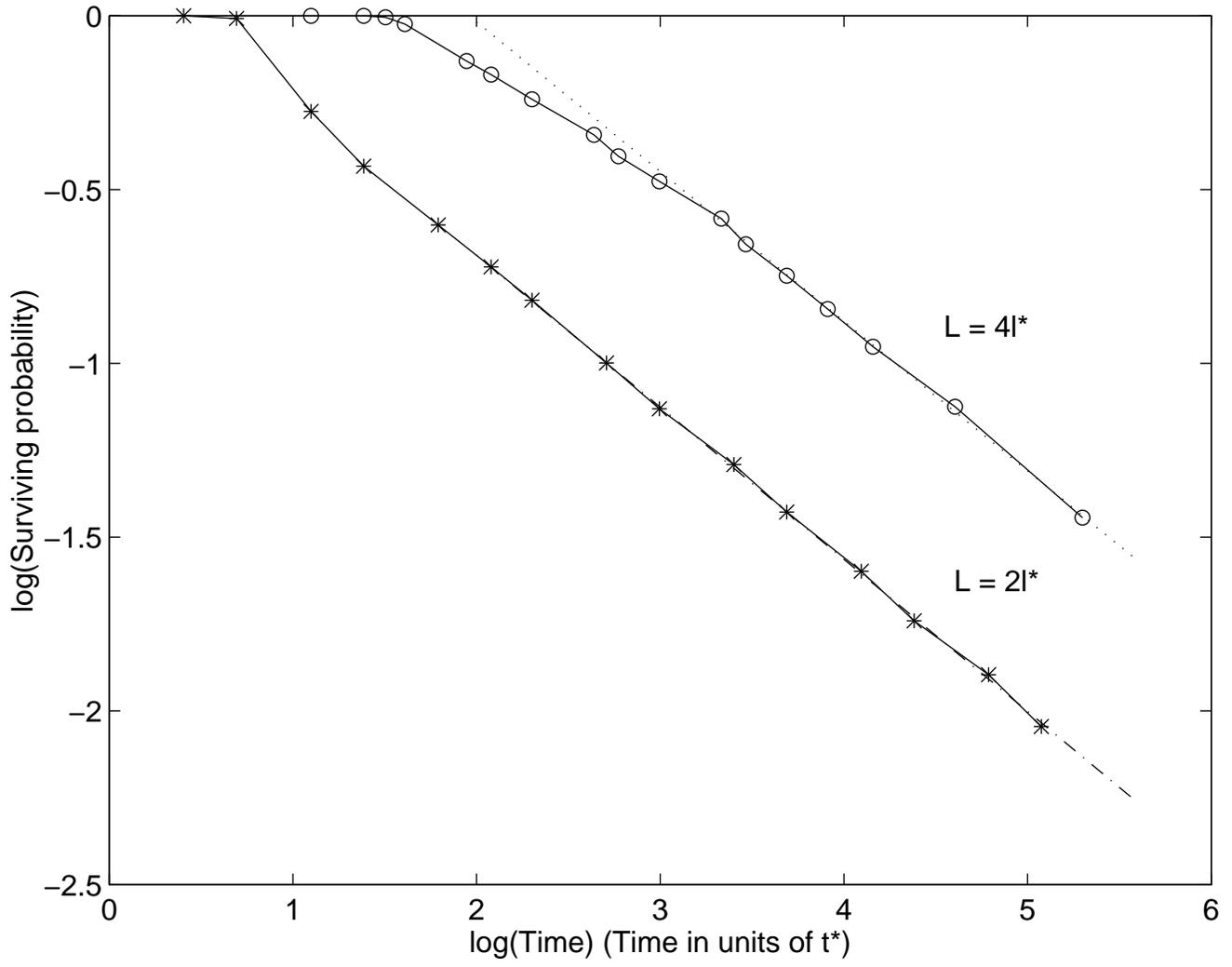
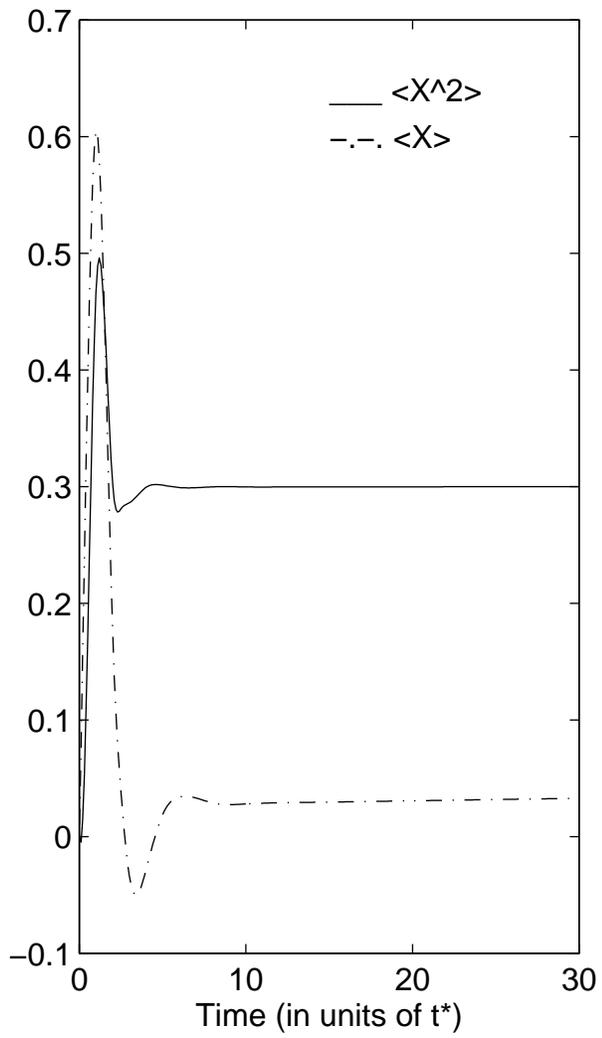
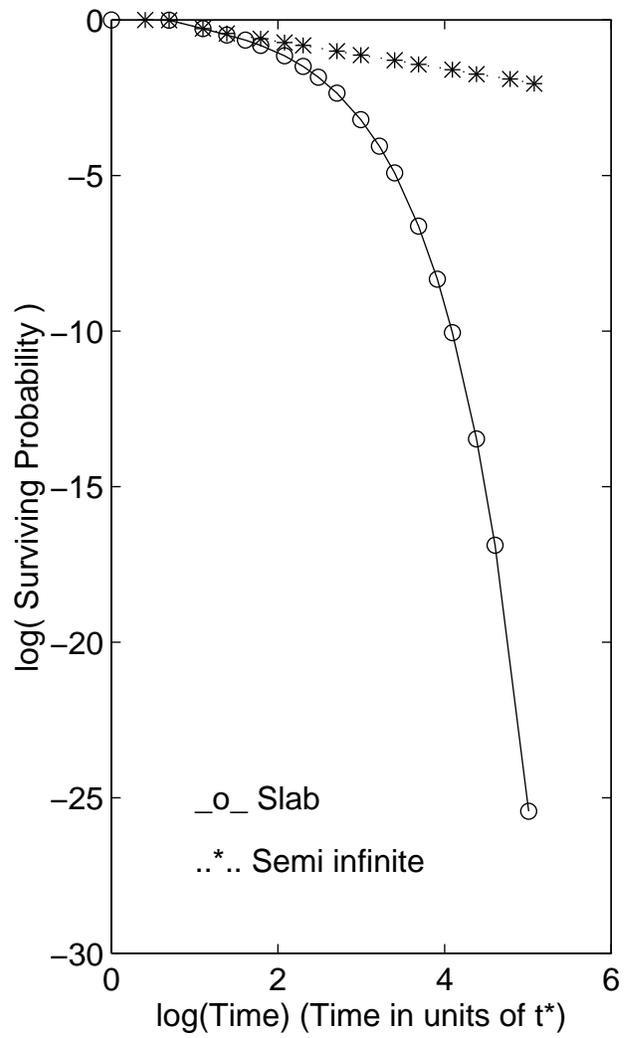
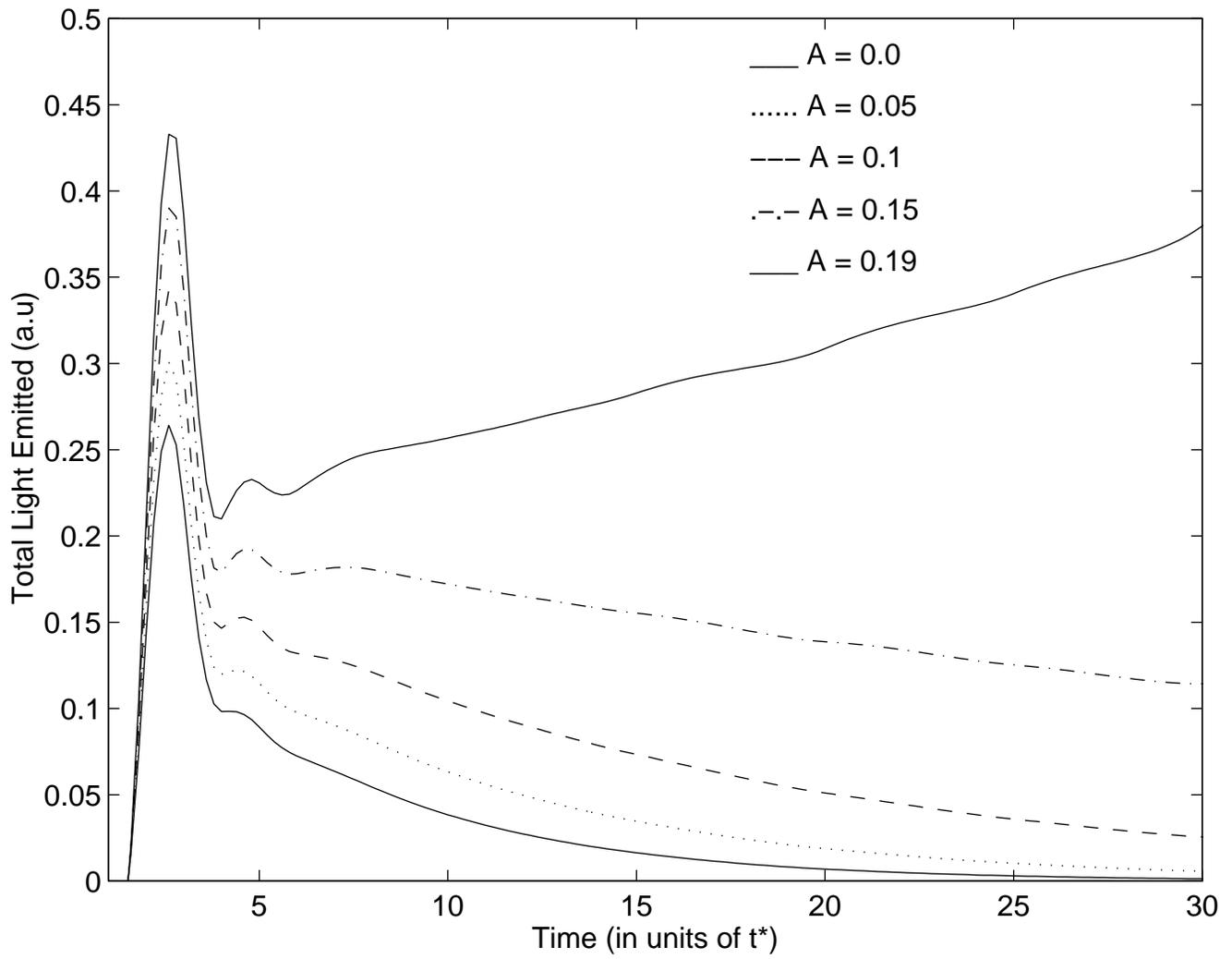
\end{document}